\def\year{2020}\relax
\documentclass[letterpaper]{article}
\usepackage{aaai20}
\usepackage{times}
\usepackage{helvet}
\usepackage{courier}
\usepackage{graphicx}
\usepackage{multirow}
\usepackage{epstopdf}
\usepackage{comment}
\usepackage{caption}
\usepackage{subcaption}
\usepackage{CJKutf8}
\usepackage[ruled]{algorithm2e}
\usepackage{booktabs}
\usepackage{amsmath}
\usepackage{float}
\usepackage[section]{placeins}
\usepackage{amsfonts,amssymb}
\usepackage{color}
\begin{document}

\title{Beyond Similarity: Relation Embedding with Dual Attentions \\for Item-based Recommendation}
\author{Liang Zhang$^{1}$, Guannan Liu$^{1}$, Junjie Wu$^{1}$\\
	$^{1}$Beihang University, \{zhangliang2017, liugn, wujj\}@buaa.edu.cn\\
	\\
}

\maketitle

\begin{abstract}
	Given the effectiveness and ease of use, Item-based Collaborative Filtering (ICF) methods have been broadly used in industry in recent years. 
	The key of ICF lies in the \emph{similarity} measurement between items, which however is a coarse-grained numerical value that can hardly capture users' fine-grained preferences toward different latent aspects of items from a representation learning perspective. In this paper, we propose a model called REDA (latent Relation Embedding with Dual Attentions) to address this challenge. REDA is essentially a deep learning based recommendation method that employs an \emph{item relation embedding} scheme through a neural network structure for inter-item relations representation. A \emph{relational user embedding} is then proposed by aggregating the relation embeddings between all purchased items of a user, which not only better characterizes user preferences but also alleviates the data sparsity problem. Moreover, to capture valid meta-knowledge that reflects users' desired latent aspects and meanwhile suppress their explosive growth towards overfitting, we further propose a \emph{dual attentions} mechanism, including a memory attention and a weight attention. A relation-wise optimization method is finally developed for model inference by constructing a personalized ranking loss for item relations. Extensive experiments are implemented on real-world datasets and the proposed model is shown to greatly outperform state-of-the-art methods, especially when the data is sparse. 
\end{abstract}

\section{Introduction}
Recommender system has become an indispensable tool in e-commerce sites for enabling users to discover their preferred items more efficiently. Numerous recommendation methods have been proposed to improve recommendation accuracy by inferring users' preferences through their purchasing records. Owing to the advantageous interpretability and effectiveness, Item-based Collaborative Filtering (ICF) methods stand out among all methods and have been widely favored by industrial applications~\cite{liu2017related,smith2017two} and become a focal point of study in recent years. Rather than adopting a handcrafted similarity measurement in traditional ICF methods, SLIM~\cite{ning2011slim} and several of its variants such as FISM~\cite{kabbur2013fism} have been proposed to learn similarity between items by taking inner dots between the latent representations of items.

In reality, however, two items may be deemed ``similar'' due to different latent factors ({\it i.e.}, relation aspects), and users may only be driven by several desired aspects. For example, two ``similar'' songs may share the same genre and singer, but this ``similarity'' could be interpreted differently for two fans; that is, one fan may be attributed to the same genre while the other may be driven by the preferred singer. From this perspective, an overall similarity metric in traditional ICF methods does lack adequate semantics to capture fine-grained user preferences toward specific aspects of items, which could generate ineffective recommendations, particularly when users have only a few purchased items without explicit intentions. This indeed motivates our study in this paper, which aims to replace item-item similarity in ICF with inter-item relations representation such that a user's preferences can be better captured by the multiple latent aspects of inter-item relations.

Therefore, in this paper, we propose a deep learning based model called REDA (latent Relation Embedding with Dual Attentions) for item-based recommendation. REDA represents each item pair in the purchased records as an \emph{item relation embedding} through a neural network structure. Then, instead of representing a user directly by an embedding vector, a \emph{relational user embedding} scheme is proposed by aggregating the item relation embeddings obtained from the whole purchasing records of a user to reflect his/her preferences in different latent aspects. Specifically, in order to construct relational user embedding, we start from decomposing all the items into multiple embeddings, with each representing a particular latent aspect. Since each embedding may represent one latent aspect, we can further exploit a crossing strategy to account for intensive interactions between different aspects of item pairs. However, not all the latent aspects of item relations would contribute equally to revealing users' preferences and too many parameters from the decomposed embedding may degrade the effectiveness in learning the model. Thus, we further propose a \emph{dual attentions} mechanism to refine the item relations in restricted relation space and avoid overfitting. We exploit a \emph{memory attention} to store the meta-knowledge that can best depict the desired latent aspects, and meanwhile a \emph{weight attention} mechanism to weight all the interacted embedding vectors. These finally yield the architecture of REDA.

Moreover, based on the architecture of item relation embedding, a relation-wise optimization method is further developed by constructing a personalized ranking loss for item relations, under the assumption that users generally maintain stable preferences toward several latent aspects, and then negative item relations can be sampled to construct the relation-wise ranking loss. Extensive experiments have been implemented on real-world datasets, and the results demonstrated that the proposed model can outperform the state-of-the-art methods, particularly on sparse data where users only have a small number of purchasing records. Additionally, we also employ a case study to show what have been learned from the latent relation embeddings in correspondence with the auxiliary information of items.

\section{Related Work}

\textbf{Item-based Collaborative Filtering.} Item-based CF methods are usually based on the principle that users prefer similar items. In this case, traditional models often predefine some similarity measures such as cosine similarity and Pearson coefficien. However, such heuristic similarity measurement lacks optimization tailored for different datasets. Recently, Ning et al. has proposed a method SLIM~\cite{ning2011slim} which learns item similarity directly from data. Afterwards, Kabbur et al. further proposes FISM~\cite{kabbur2013fism} to explore the low-rank property of the learned similarity matrix to handle data sparsity problem. While FISM is shown to outperform recommendation approaches, it has the limitation in estimating only a single global metric for all users. To that end, GLSLIM~\cite{christakopoulou2016local} clusters the users and estimates an independent SLIM model for every user subset, whereas the number of clusters is difficult to determine, and thus may yield suboptimal results. 

\textbf{Deep Learning in Recommendation.} Recently employing deep learning to help improve recommendation performance has become prevailing in the research field. NeuMF~\cite{he2017neural} addresses implicit feedback by jointly learning a matrix factorization and a feedforward neural network. Following this research line, NNCF~\cite{bai2017neural} is further proposed as a variant that takes user neighbors and item neighbors as inputs. Liang et al. extends NeuMF by substituting MLP with auto-encoder architecture\cite{liang2018variational}. It is worth noting that all these methods focus on the user-item latent space and overlook the correlations in the item-item latent space. Besides the more closely related domain of collaborative filtering on implicit feedback, CNN is often used to capture localized item feature representations of images~\cite{he2016vbpr} and text~\cite{kim2016convolutional}. The sequential nature of RNN also provides desirable properties for time aware~\cite{wu2017recurrent} and session-based recommendation systems~\cite{hidasi2015session}. However, this auxiliary information is not always available, which limits their applications in real-world recommendation systems.

\textbf{Memory Augmented Neural Networks.} Memory augmented neural networks have shown significant success in NLP research areas~\cite{sukhbaatar2015end,kumar2016ask,miller2016key} recently. The key insight is using external memory components to assist the deep neural networks in remembering and storing meta-knowledge. Motivated by this architecture, RUM~\cite{chen2018sequential} replaces common RNN with memory network to address sequential recommendation. Gamenet~\cite{shang2019gamenet} uses memory component to fuse multi-model graphs as memory bank to facilitate medication combination recommendation. As for top-\emph{n} recommendation tasks, CMN~\cite{ebesu2018collaborative} introduces the memory module to store user embeddings as a nearest neighborhood model identifying similar users. LRML~\cite{tay2018latent} exploits the key-value memory neural network to model relational translations between user and item vectors. We follow this trend to enhance our model by devising a memory module to model the relations between two arbitrary items.

\section{Methodology}

\begin{figure}[t!]
	\centering
	\includegraphics[width=0.48\textwidth]{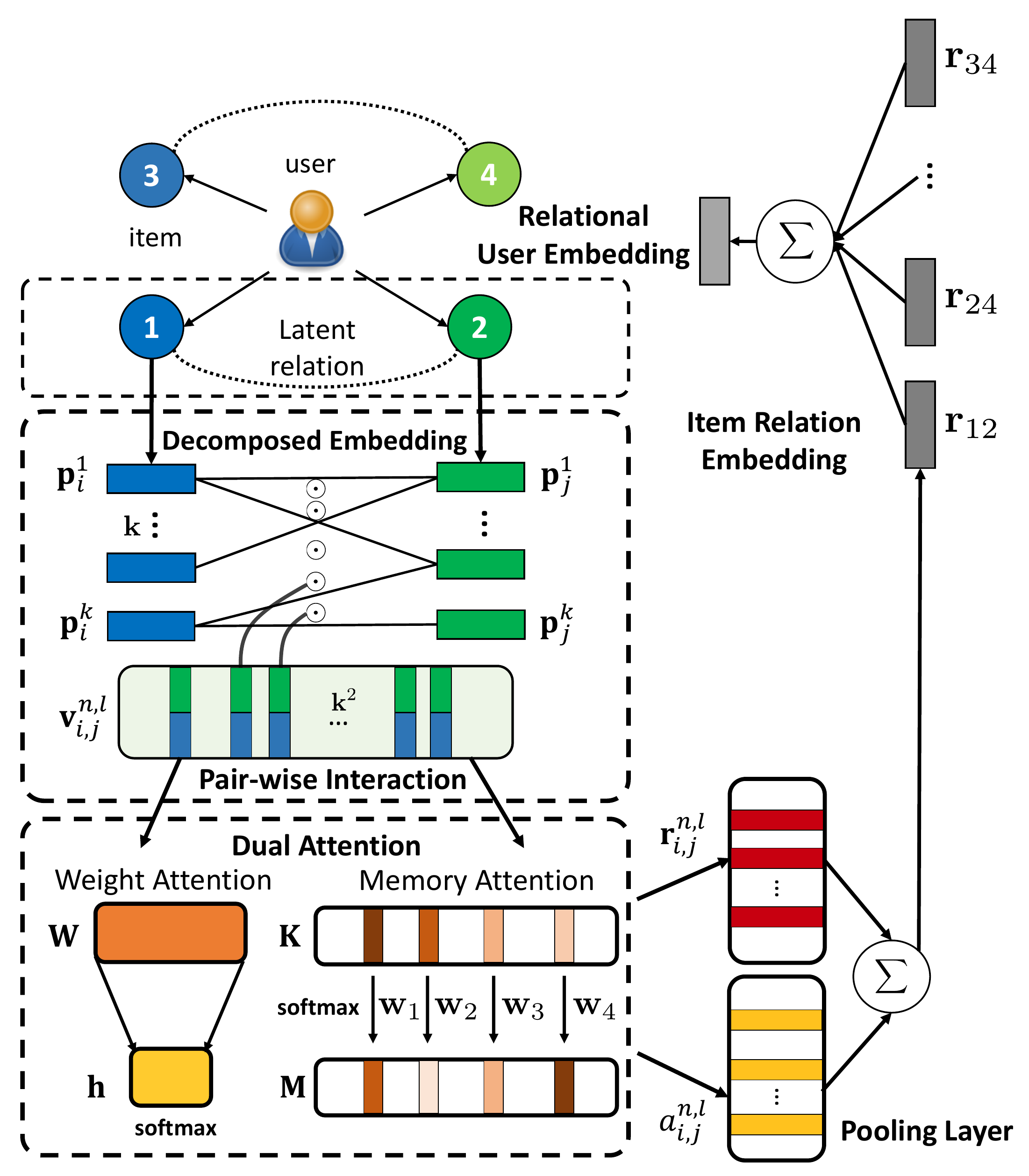}
	\caption{The architecture of REDA.}
	\label{fig:framework}
\end{figure}

The core idea of item-based CF method is to predict a user's rating on a target item based on the similarity between this item and all the previously rated items of the user, which is generally in the following form according to~\cite{kabbur2013fism}.
\begin{equation}
\hat{y_{u,i}} = \mathbf{p}_i^\top(\frac{1}{|\mathcal{R}_u^+|^\alpha} \sum_{j \in \mathcal{R}_u^+\backslash\{i\}}\mathbf{q}_j),
\label{eq:fism}
\end{equation}
where $\mathbf{p}_i$ and $\mathbf{q}_j$ denote the embedding vector for item $i$ and $j$ respectively, and $\mathcal{R}_u^+$ denotes the set of items that have been purchased by $u$. In this approach, the similarity is obtained through the inner product between $\mathbf{p}_i$ and $\mathbf{q}_j$. 

However, if we take a subtle look at the similarity between item pairs, one prominent problem would naturally arise, that is, an overall similarity cannot provide adequate semantics to disclose specific aspects for the item pair.

Obviously, a simple inner product cannot capture such complex relations between items, and meanwhile each user may only desire a partial latent aspects for items. To address this concern, one of the feasible solutions might be to decompose the similarity into more fine-grained item relations that could account for multiple aspects, and users' preferences toward specific aspects can then be revealed.

Therefore, we propose a novel model \emph{latent Relation Embedding with Dual Attentions} (REDA) to handle the above limitations regarding the item similarity. The overview of the model architecture is shown in Fig.~\ref{fig:framework}. We then introduce each module of the model in detail as follows.

\subsection{Item Relation Embedding}
Since an overall similarity cannot adequately capture users' fine-grained preferences, we seek to better represent the relation between a pair of items so as to decompose the overall similarity measurement. Specifically, we extend the general idea of Item-based CF method that a user may prefer items similar to those in their purchasing records, by adding a more subtle assumption that the ``similarity'' should be owing to several particular aspects. In this regard, a pair of items that have been bought by a same user can generate a relation, with emphasis on several latent aspects. Notably, a paired item may appear in different users' records, but the corresponding driving aspects for the item relation may differ according to the whole purchasing records of a user. For example, two songs may share the same singer and genre, but two different users may be keen on different aspects, which can only be revealed by probing more item relations of the users respectively.
Thus, in order to formulate more comprehensive item relation embeddings, we need to firstly decompose a user's purchasing records to paired items.

Then, we employ a special decomposed embedding layer to represent each item by multiple embedding vectors to denote various aspects of the item. Specifically, each item $i$ is represented by a $k$ embedding vector $\mathbf{p}_{i}^1,\cdots,\mathbf{p}_{i}^k$.

\paragraph{Pair-wise Interaction Layer}
For a given item pair $i$ and $j$, we can obtain $\mathbf{p}_{i}$ and $\mathbf{p}_{j} \in \mathbb{R}^{k \times d}$ through the decomposed embeddings, where $d$ is the dimensionality of item embeddings. However, the relations between each item pair may not be exactly captured though one-to-one mapping between the corresponding embedding vectors. Inspired by previous methods \cite{rendle2010factorization,wang2017deep} that considers the interaction across features, we augment the $k$ embedding vectors to $k^2$ interacted vectors $\mathbf{v}_{i,j}$ by crossing the embeddings representing different latent aspects, where $\mathbf{v}_{i,j}^{n,l}$ is the element-wise product of $\mathbf{p}_{i}^n$ and $\mathbf{p}_{j}^l$. Formally, the output of this layer can be represented as follows,

\begin{equation}
f_{PI}(i,j) = {\{\mathbf{p}_{i}^n \odot \mathbf{p}_{j}^l\}}_{1 \leq n \leq k, 1 \leq l \leq k},
\end{equation}
where $\odot$ denotes the element-wise product of two vectors. By defining this layer, we can model the interactions between items in different latent aspects.

\paragraph{Dual Attentions Layer}
Given the decomposed embeddings with pair-wise interactions for each item pair, we propose a dual attentions layer to derive a more delicate relation embedding.
On one hand, we need to reason which aspects would these latent interacted embedding vectors actually implicate, in which we exploit an \emph{memory attention} mechanism to reason and store valid meta-knowledge. On the other, not all the interacted embeddings can contribute to the relation embedding, and thus we add another \emph{weight attention} to account for the importance of different interactions.

\textbf{Memory Attention}: One assumption we have made for the item relations is that they can indicate several latent aspects between the items, \emph{e.g.}, the genre, the singer, the price, and other attributed aspects of the items, which can generally be regarded as prior knowledge for the relations. Therefore, the core of this framework is to reason the relation aspects of two arbitrary items from the prior knowledge, given the interacted embeddings. This procedure is indeed similar to answering a complex question, where the interacted embedding vectors can be treated as query vector while the output is the answer, namely the corresponding item relation embedding.
Following this idea, we introduce a key-value structured memory, which is often used in question answering~\cite{miller2016key}, to generate item relation embedding. Through the memory attention module, useful meta-knowledge towards a particular pair of interacted embedding vector can be captured. Specifically, the memory matrix is represented as $\mathbf{M} \in \mathbb{R}^{m \times d}$ where $m$ is the number of memory slices. Similarly, the key matrix is represented as $\mathbf{K} \in \mathbb{R}^{m \times d}$. For a specific interacted vector $\mathbf{v}_{i,j}^{n,l}$ obtained in pair-wise interaction layer, we generate corresponding item relation embedding vector as follows. 

\begin{equation}
a_{t} = {\mathbf{v}_{i,j}^{n,l}}  \mathbf{k}_t ^\top,
\end{equation}
\begin{equation}
\tilde{a_t}=\text{softmax}(a_{t}) = \frac{\exp(a_{t})}{\sum_{t^\prime} \exp(a_{t^\prime})},
\end{equation}
\begin{equation}
\mathbf{r}_{i,j}^{n,l} = {\sum_{t} \tilde{a}_t \mathbf{m}_{t}},
\end{equation}
where $a_{t}$ is the normalized attention score and $\mathbf{r}_{i,j}^{n,l}$ is the generated item relation embedding vector. 

\textbf{Weight Attention}: Intuitively, there might exist some trivial interactions between items, and it is unreasonable to assign an equal weight for all interacted vectors. Therefore, we employ the general attention mechanism which has been widely used in many tasks. The idea is to allow different interactions contribute differently when compressing them to a single item relation representation. Specifically, the attention score of a specific interacted vector can be obtained by a Multi-Layer Perception (MLP).

\begin{equation}
a_{i,j}^{n,l} = \mathbf{h}^\top \text{ReLU}(\mathbf{W} \mathbf{v}_{i,j}^{n,l} + \mathbf{b}),
\end{equation}

\begin{equation}
\label{eq:attweight}
\tilde{a}_{i,j}^{n,l}=\text{softmax}(a_{i,j}^{n,l}) = \frac{\exp(a_{i,j}^{n,l})}{\sum_{n^\prime,l^\prime} \exp(a_{i,j}^{n^\prime,l^\prime})},
\end{equation}
where $\mathbf{W} \in \mathbb{R}^{d \times s}$, $\mathbf{b} \in \mathbb{R}^{s}$, $\mathbf{h} \in \mathbb{R}^{s}$ are model parameters, and $s$ denotes the hidden layer size.

\paragraph{Pooling Layer}
The output of pooling layer is a $d$ dimension vector namely $\mathbf{r}_{i,j}$, which compress all item interactions between item $i$ and $j$ in the relation space by distinguishing their importance. 

\begin{equation}
\mathbf{r}_{i,j} = \sum_{1 \leq n \leq k, 1 \leq l \leq k} \tilde{a}_{i,j}^{n,l} \mathbf{r}_{i,j}^{n,l},
\end{equation}
where $\mathbf{r}_{i,j}^{n,l}$ is the relation embedding and $a_{i,j}^{n,l}$ is the corresponding attention weight calculated by Equation~\eqref{eq:attweight}.

\subsection{Relational User Embedding}
Assuming that users would remain stable preferences toward particular aspects in choosing items, then if we probe all the possible item relations formed in their purchasing records, their preferences toward certain aspects can be distilled from the aggregated item relations. For instance, we cannot directly identify the underlying preferences for the two users listening to the same pair of songs, but if we inspect all the relations formed in their playlists respectively, we may find the songs in one playlist mostly share the same genre, while the other concentrates on a same singer.

Therefore, personalized preferences of a user $u$ on particular latent aspects can be disclosed through the item relations decomposed from their purchasing records $\mathcal{R}^+_u$. By aggregating relation embeddings between all the item pairs of each individual user, we can map users into the same relational space to model their preferences. We refer $\mathbf{z}_u$ to represent user embedding of user $u$, which can be derived by,

\begin{equation}
\mathbf{z}_u = \sum_{i,j \in \mathcal{R}_u^+, i \neq j} \mathbf{r}_{i,j}.
\end{equation}

Since each user has their own purchasing records, the aggregated item relation embeddings can concentrate their fine-grained preferences on certain latent aspects. Moreover, an item relation embedding is learned by traversing all the users' purchasing records, and thus all the possible desired latent aspects of any two co-purchased items can be captured across users. 
In view of this, a relation embedding learned from REDA is universally applicable among all the users. Thus, even when users have only a small number of purchased items, \emph{i.e.}, the purchasing data is sparse, the learned relation embedding can provide adequate information to uncover their preferences and may still ensure good recommendation performances. 

\subsection{Relation-wise Optimization}
Traditional solution for learning Item-based CF is item-wise optimization, which maximizes the difference between user-interacted items and non-interacted items. Unfortunately, we find such a standard solution does not work well in practice. Therefore, in this paper, we propose a novel relation-wise optimization method to guide the learning of REDA. An underlying assumption is that users would remain stable in their preferences for the items and therefore relations between user-interacted items tend to be more similar. Specifically, given a particular user $u$ and a user-interacted item pair $(i_t,j_t)$, we randomly sample another interacted item pair $(i_c,j_c)$ of the user, and meanwhile we also sample a non-interacted item pair $(i_n,j_n)$. Then, $\mathbf{r}_{i_c,j_c}$ should be more close in the relational space with $\mathbf{r}_{i_t,j_t}$ than that with $\mathbf{r}_{i_n,j_n}$. Following this idea, we can construct the loss function in view of item relations as follows.

\begin{equation}
\mathcal{L} = \sum_{u \in \mathcal{U}} \sum_{i_t,j_t \in \mathcal{R}_u^+} \log\sigma(\mathbf{r}_{i_c,j_c}^\top \mathbf{r}_{i_n,j_n} - \mathbf{r}_{i_c,j_c}^\top \mathbf{r}_{i_t,j_t}).
\end{equation}

Here, we name $\mathbf{r}_{i_t,j_t}$ as positive relation, $\mathbf{r}_{i_c,j_c}$ as context relation and $\mathbf{r}_{i_n,j_n}$ as negative relation. The relation-wise optimization is to maximize the distance between negative relation and context relation while minimize the distance between positive relation and context relation simultaneously.

\subsection{Recommendation Score}
Similar to traditional Item-based CF, when evaluating the recommendation score of user $u$ for a target item $i$ given the learned representations, we need to revisit the relations between $i$ and each item $j$ that has ever been purchased by $u$, which can be approximated by,

\begin{equation}
\hat{y_{u,i}} = \frac{1}{|\mathcal{R}_u^+|} \sum_{j \in \mathcal{R}_u^+\backslash\{i\}} \mathbf{z}_u^\top \mathbf{r}_{i,j}.
\label{eq:score}
\end{equation}

Compared to Equation~\eqref{eq:fism}, the main difference is to replace the global similarity metric $\mathbf{p}_i^\top \mathbf{q}_j$ with personalized preferences on relation aspects $\mathbf{z}_u^\top \mathbf{r}_{i,j}$. When generating recommendations for $u$, we simply need to rank candidate items according to the recommendation score and select the ones with highest scores as the items to be recommended.

\section{Experiment}
In this section, we first evaluate the model performance in terms of recommendation accuracy, and analyze the influences of each component respectively (\emph{i.e.}, the pair-wise interaction layer and memory attention layer). Then we conduct experiments to demonstrate the robustness and data sparsity tolerance of our model. Finally, we give a short parameter sensitivity discussion and a case study to explain what the latent relation embeddings represent.

\subsection{Experimental Setup}

\paragraph{Data Sets.}  We evaluate our method on four real-world  datasets with diverse sizes and interaction intensity, which are all publicly available datasets to validate recommendation algorithms including Delicious\footnote{http://www.delicious.com}, LastFM\footnote{http://last.fm}, Amazon-Music\footnote{https://www.amazon.com}~\cite{mcauley2015image}, and Goodreads-Poetry\footnote{https://www.goodreads.com}~\cite{wan2018item}.
In this paper, we treat rating larger than 3 as positive feedback, and only retain users having more than 5 actions. The detailed statistics of all the datasets are reported in Table~\ref{tab:data}.

\begin{table}[!t]
	\centering
	\normalsize
	\caption{Data statistics.}
	\label{tab:data}
	\resizebox{0.9\linewidth}{!}{
		\begin{tabular}{lcccc}
			\toprule
			& \# users & \# items & \# actions & \% Density \\
			\midrule
			Delicious & 1779 & 68326 & 102822 & 0.08 \\ 
			LastFM & 1877 & 17543 & 90924 & 0.27 \\ 
			Amazon-Music & 19847 & 95426 & 181124 & 0.01 \\ 
			Goodreads-Poetry & 45096 & 31604 & 540546 & 0.04 \\ \bottomrule
	\end{tabular}}
\end{table}

\paragraph{Baseline Methods.} The following SOTA methods are applied as baselines in our experiments.

\textbf{NMF}~\cite{Paatero2010Positive}: NMF is a widely used collaborative filtering approach, which factorizes the interaction binary matrix.

\textbf{BPR-MF}~\cite{rendle2009bpr}: BPR-MF is a well-known top-\emph{n} recommendation method, which uses the Bayesian personalized ranking optimization criterion on factorization.

\textbf{FISM}~\cite{kabbur2013fism}: FISM is a state-of-the-art item-based CF method which learns global item similarities from user-item interactions.

\textbf{NeuMF}~\cite{he2017neural}: NeuMF is a unified framework to combine MF with MLP. NeuMF concatenates the output of MF and MLP, and uses a regression layer to predict user item rating.

\textbf{CML}~\cite{hsieh2017collaborative}: CML is a state-of-the-art metric based CF method,  which learns a joint metric space by encoding both users' preferences and similarity between users and items.

\textbf{LRML}~\cite{tay2018latent}: LRML extends CML by explicitly modeling relational translations between user and item vectors.

\begin{table*}[t!]
	\centering
	\caption{Experimental results on four real-world datasets.}
	\label{tab:rec_result}
	\resizebox{0.85\textwidth}{!}{
		\begin{tabular}{@{}llccccccccc@{}}
			\toprule
			\multirow{2}{*}{Method} & \multicolumn{2}{c}{{\it Delicious}} & \multicolumn{2}{c}{{\it LastFM}} & \multicolumn{2}{c}{{\it Amazon-Music}} & \multicolumn{2}{c}{{\it Goodreads-Poetry}} \\
			\cmidrule{2-9}
			& HR@10 & nDCG@10 & HR@10 & nDCG@10 & HR@10 & nDCG@10 & HR@10 & nDCG@10 \\ \midrule
			NMF & 0.4942 & 0.2960 & \underline{0.7766} & 0.4996 & 0.6062 & \underline{0.4358} & 0.6904 & 0.5541 \\
			BPR-MF & 0.4665 & 0.2753 & 0.6138 & 0.3444 & 0.5268 & 0.3574 & 0.7756 & 0.6161 \\
			FISM & 0.3966 & 0.2547 & 0.7273 & 0.5145 & \underline{0.6173} & 0.4078 & 0.7843 & 0.6365 \\
			CML & 0.4825 & 0.3714 & 0.7039 & 0.4729  & 0.5278 & 0.3795 & 0.8251 & \underline{0.7046} \\
			LRML & 0.4018 & 0.2932 & 0.6218 & 0.4064 & 0.4559 & 0.2758 & 0.7459 & 0.5816 \\
			NeuMF & \underline{0.5423} & \underline{0.3873} & 0.7715 & \underline{0.5186} & 0.5936 & 0.4154 & \underline{0.8322} & 0.7015\\
			\cmidrule{1-9}
			REDA & \textbf{0.5754} & \textbf{0.4320} & \textbf{0.7944} & \textbf{0.5288} & \textbf{0.6738} & \textbf{0.4952} & \textbf{0.8862} & \textbf{0.7486} \\ \bottomrule
	\end{tabular}}
\end{table*}

\paragraph{Experimental Settings.} In this section, we adopt the leave-one-out evaluation strategy, that is, for each user, we hold-out one purchased item as test set and the remaining is used for training. Since it is too time-consuming to rank all the items for every user during evaluation, we follow the experimental settings in~\cite{he2017neural} which randomly samples 100 negative items of each user and rank the recommendation score among the 100 items. We evaluate the performances of REDA based on HR@N and nDCG@N given the top-N ranked items. 

The parameter settings of different methods are stated as follows. For our method, we set the mini-batch size, the learning rate of the Adam~\cite{kingma2014adam}, the hidden layer size $s$, the embedding size $d$ to be 2000, 0.001, 10 and 128 respectively. The number of memory slices $m$ is tuned amongst \{5,10,20,30,40,50\}, and the number of $k$ for pair-wise interaction layer is tuned amongst \{2,3,4,5,10\}. As for the baseline methods, we apply default parameters except for the embedding size $d$, which is fixed to be 128 for all the methods. More detailed discussions about parameter sensitivity of our model will be given in the following subsection.

\subsection{Performance Comparisons}
We show the recommendation performance of different methods in Table~\ref{tab:rec_result}, where the best performance is in boldface and second best is underlined. We can find that the proposed model is consistently better than all the baselines as measured by HR and nDCG in all the datasets, while the second best is relatively unstable, showing that the model REDA is more robust against various settings. 
Moreover, REDA achieves the greatest improvement on {\it Amazon-Music} with the most sparse ratings among the four datasets, which indicate that the model is capable of handling data sparsity issues and more results will be shown on this. NeuMF generally performs better than other baseline methods in many cases, however it does not perform well on {\it Amazon-Music}.
It is worth mentioning that REDA is much better than FISM, which may be possibly due to that the decomposed similarity through item relation embedding can capture fine-grained user preferences. 
Among the baselines, the performance of NeuMF is often better than other methods in most cases. In contrast, CML seems to perform reasonably well only on {\it Goodreads-Poetry}, the relatively larger and denser dataset, which may indicate that CML cannot handle data sparsity issues.
In addition, the performance of LRML is not satisfactory on all datasets though the relations between items and users are explicitly modeled with a memory network. A possible reason might be that the relations are too complex to be effectively modeled with limited memory slices. 
While in our case, item relations are relatively stable and limited, hence may be explained by several aspects.

\subsection{Model Discussion}
\begin{figure}[t!]
	\centering
	\begin{subfigure}[b]{0.22\textwidth}
		\includegraphics[width=\textwidth]{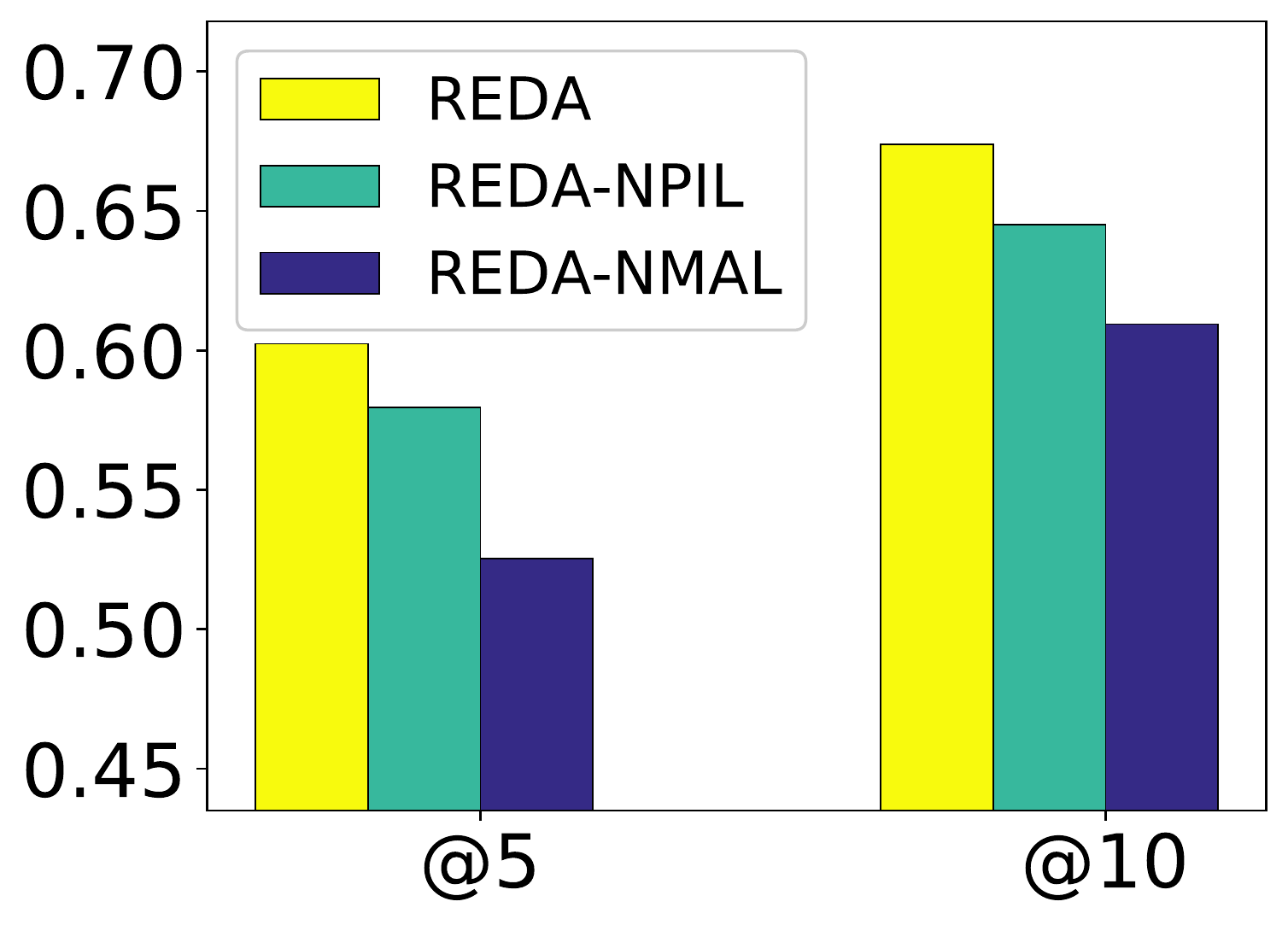}
		\caption{HR}
		\label{fig:ex_2_hr}
	\end{subfigure}
	\begin{subfigure}[b]{0.22\textwidth}
		\includegraphics[width=\textwidth]{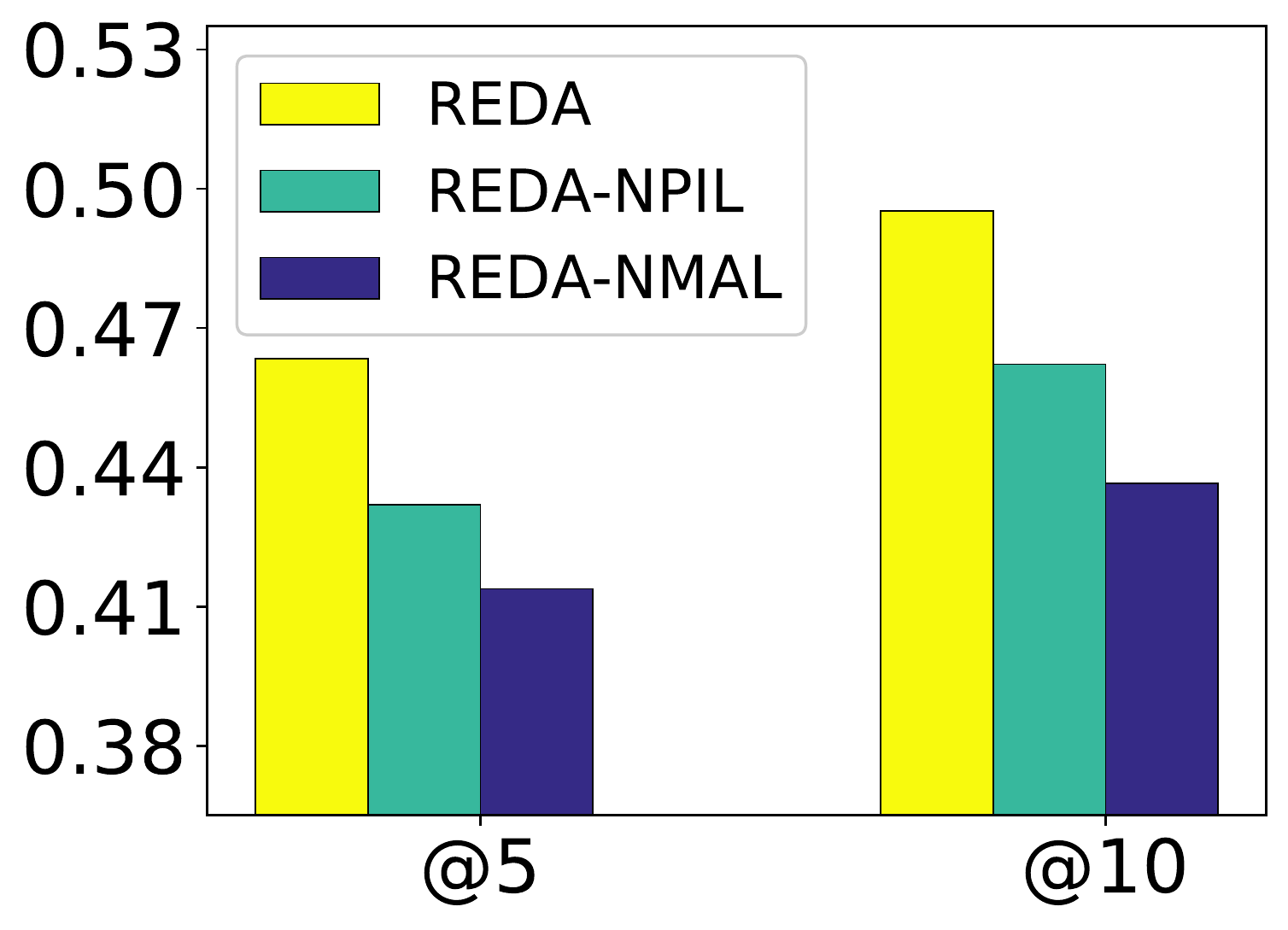}
		\caption{nDCG}
		\label{fig:ex_2_ndcg}
	\end{subfigure}
	\caption{Module influences on model performances.}
	\label{fig:model_discussion}
\end{figure}

We further inspect how each module of the model affects the performances in Figure~\ref{fig:model_discussion}. Specifically, by removing the pair-wise interaction layer and memory attention layer, we can obtain two simple versions of our model, namely REDA-NPIL and REDA-NMAL respectively. Here, removing the pair-wise interaction layer means we just have one embedding for each item and one interacted vector for each item pair. We observe that the complete REDA improves REDA-NMAL for at least 8\% in both metrics, showing that memory attention layer is a key component and greatly affects model performances. Moreover, REDA is consistently better than REDA-NPIL in all the settings, meaning that multiple item embeddings can indeed capture item combinations in different latent aspects more precisely.

\subsection{Data Sparsity}
\begin{figure}[t!]
	\centering
	\begin{subfigure}[b]{0.22\textwidth}
		\includegraphics[width=\textwidth]{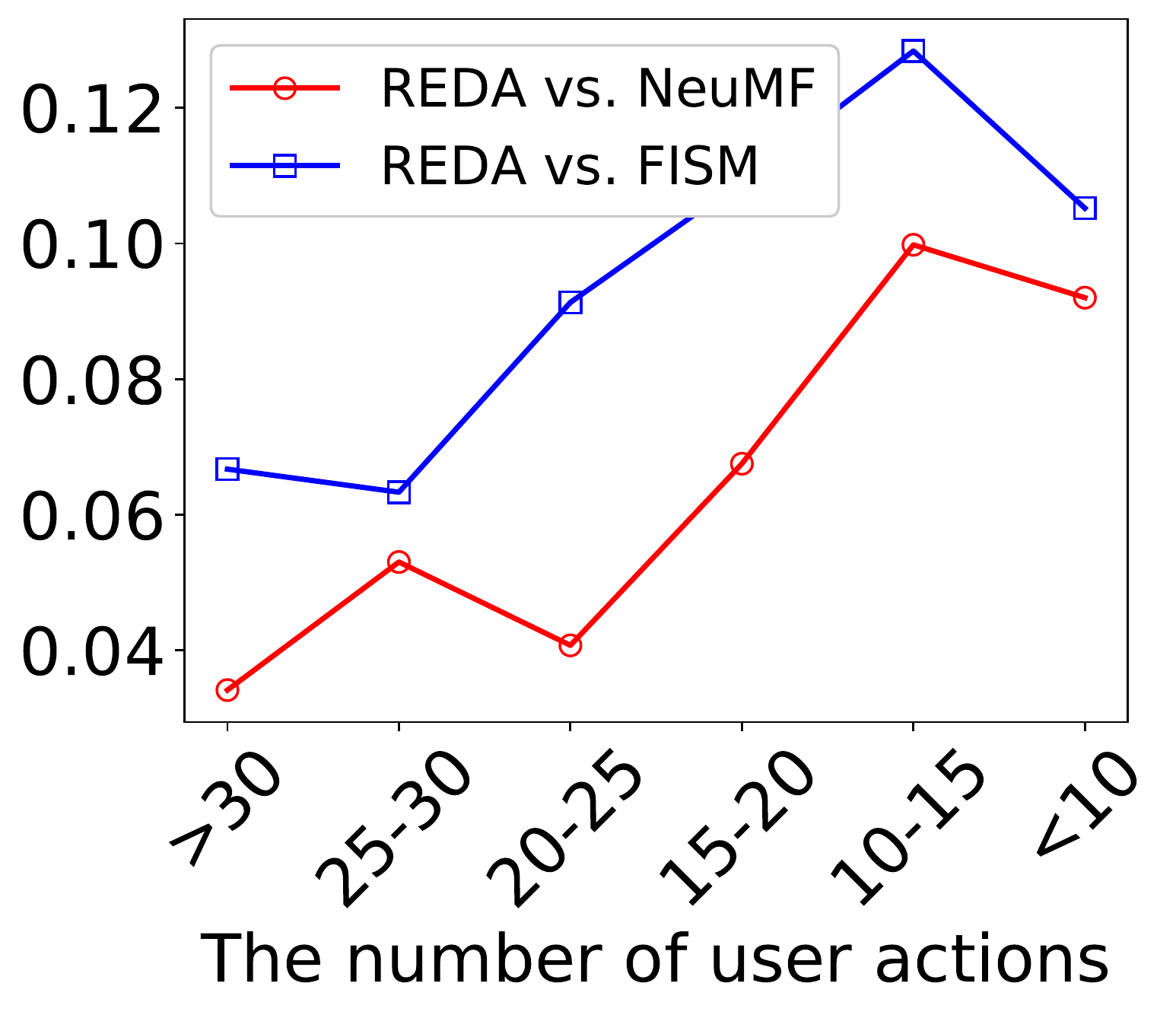}
		\caption{HR@10}
		\label{fig:ex_4_HR}
	\end{subfigure}
	\begin{subfigure}[b]{0.22\textwidth}
		\includegraphics[width=\textwidth]{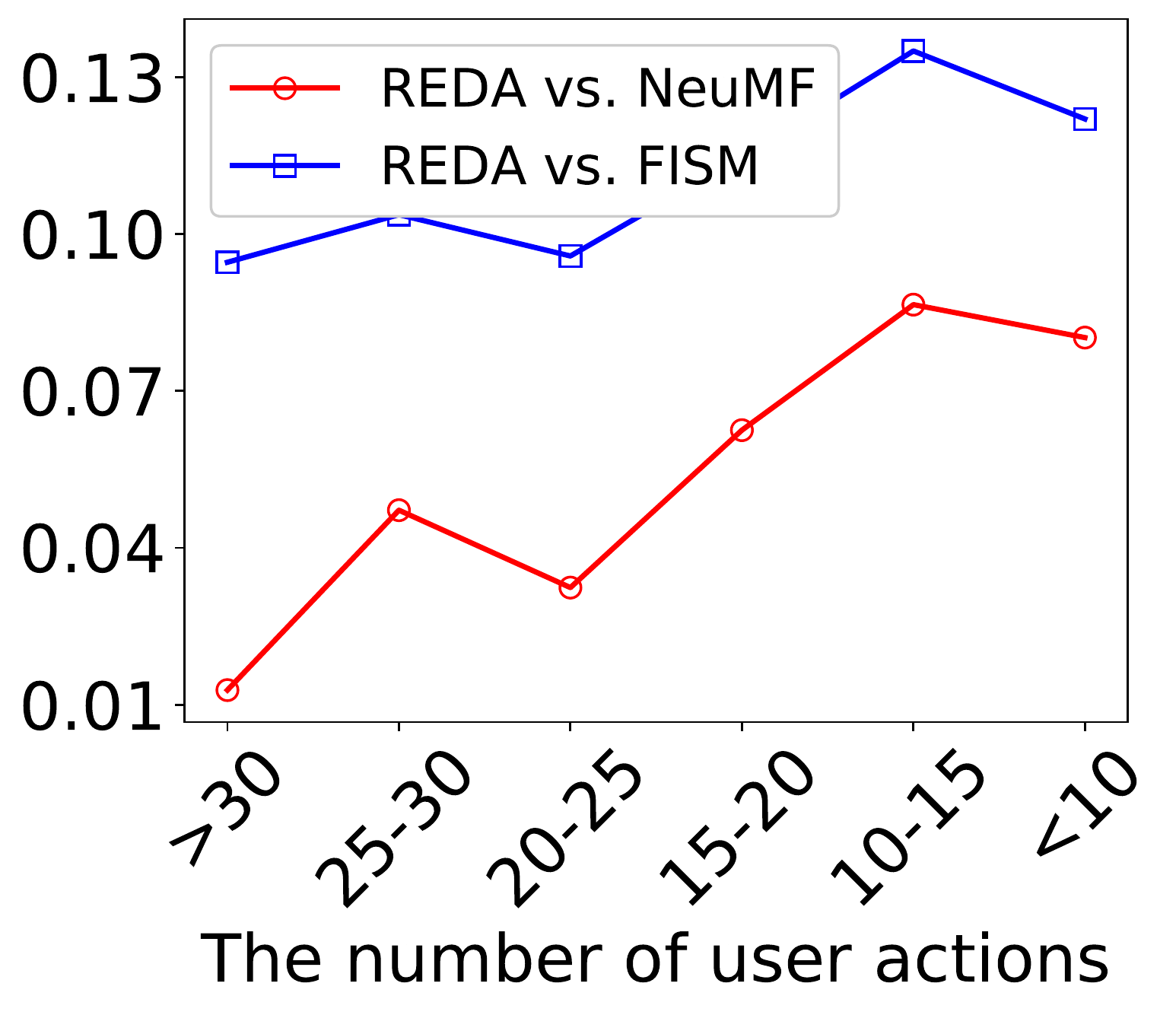}
		\caption{nDCG@10}
		\label{fig:ex_4_ndcg}
	\end{subfigure}
	\caption{ Improvement achieved by REDA over other methods as data sparsity increases}
	\label{fig:sparsity}
\end{figure}

In real-world recommendations, data sparsity is one big challenge because most users only have sparse purchasing records. In this section,
we proceed to further explore whether the proposed model can address this challenge. To that end, we conduct comparative experiments on several focal groups of users under different sparsity settings.
Specifically, we still adopt the leave-one-out evaluation strategy, and group users according to the number of their actions to derive the recommendation performances. Figure~\ref{fig:sparsity} shows the experimental results on different groups of users, where we present the actual improvement of REDA over baseline methods NeuMF and FISM. It is obvious that REDA outperforms both NeuMF and FISM increasingly as sparsity grows, \emph{i.e.}, the number of actions decreases. NeuMF can achieve comparable performances as REDA when the number of user actions larger than 30, but will drop rapidly when data becomes sparse. 
This result well demonstrates the effectiveness of REDA in handling sparsity issues due to the intensive learned relation embeddings. 

\subsection{Model Robustness}
\begin{figure}[t!]
	\centering
	\begin{subfigure}[b]{0.22\textwidth}
		\includegraphics[width=\textwidth]{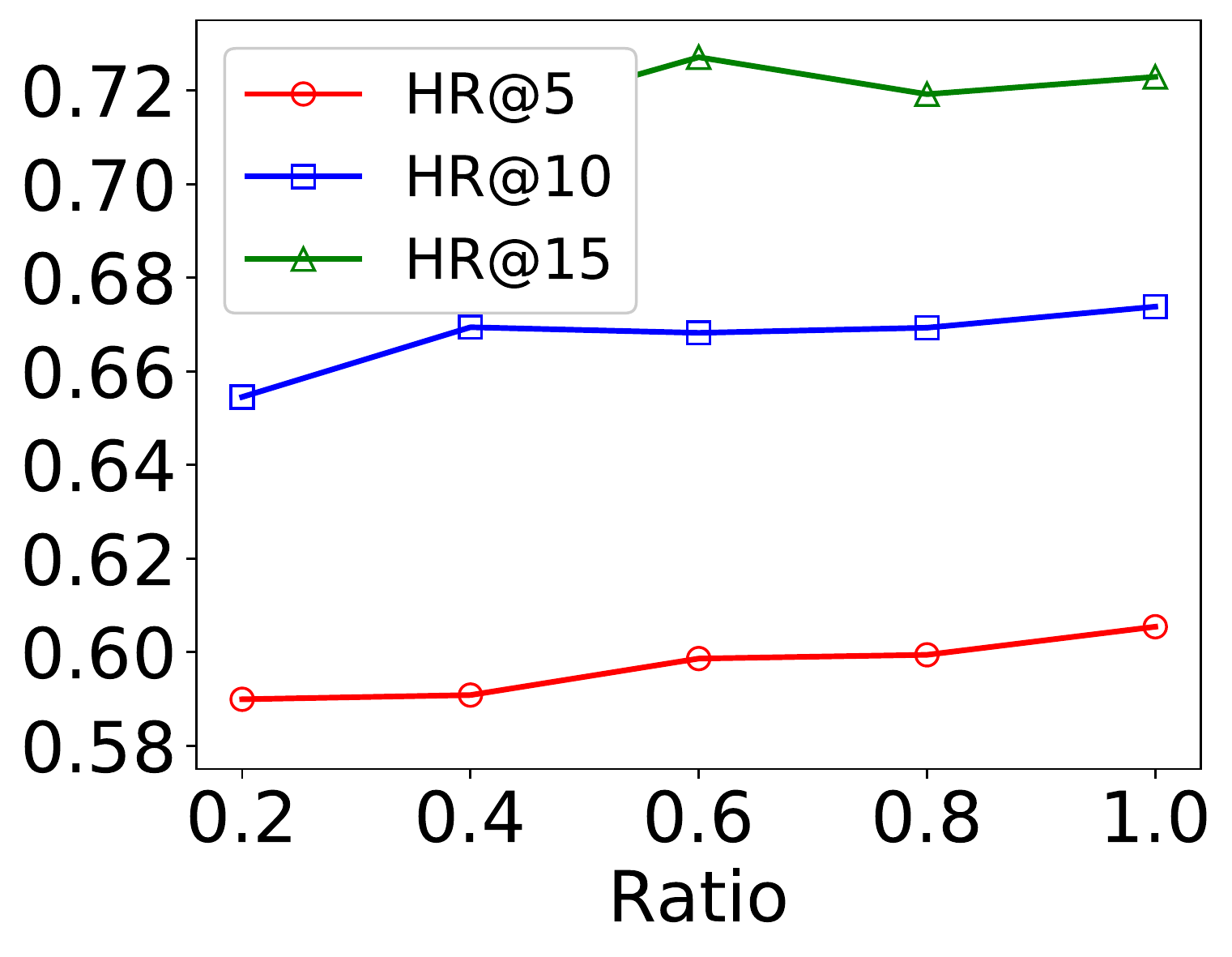}
		\caption{HR}
		\label{fig:ex_3_HR}
	\end{subfigure}
	\begin{subfigure}[b]{0.22\textwidth}
		\includegraphics[width=\textwidth]{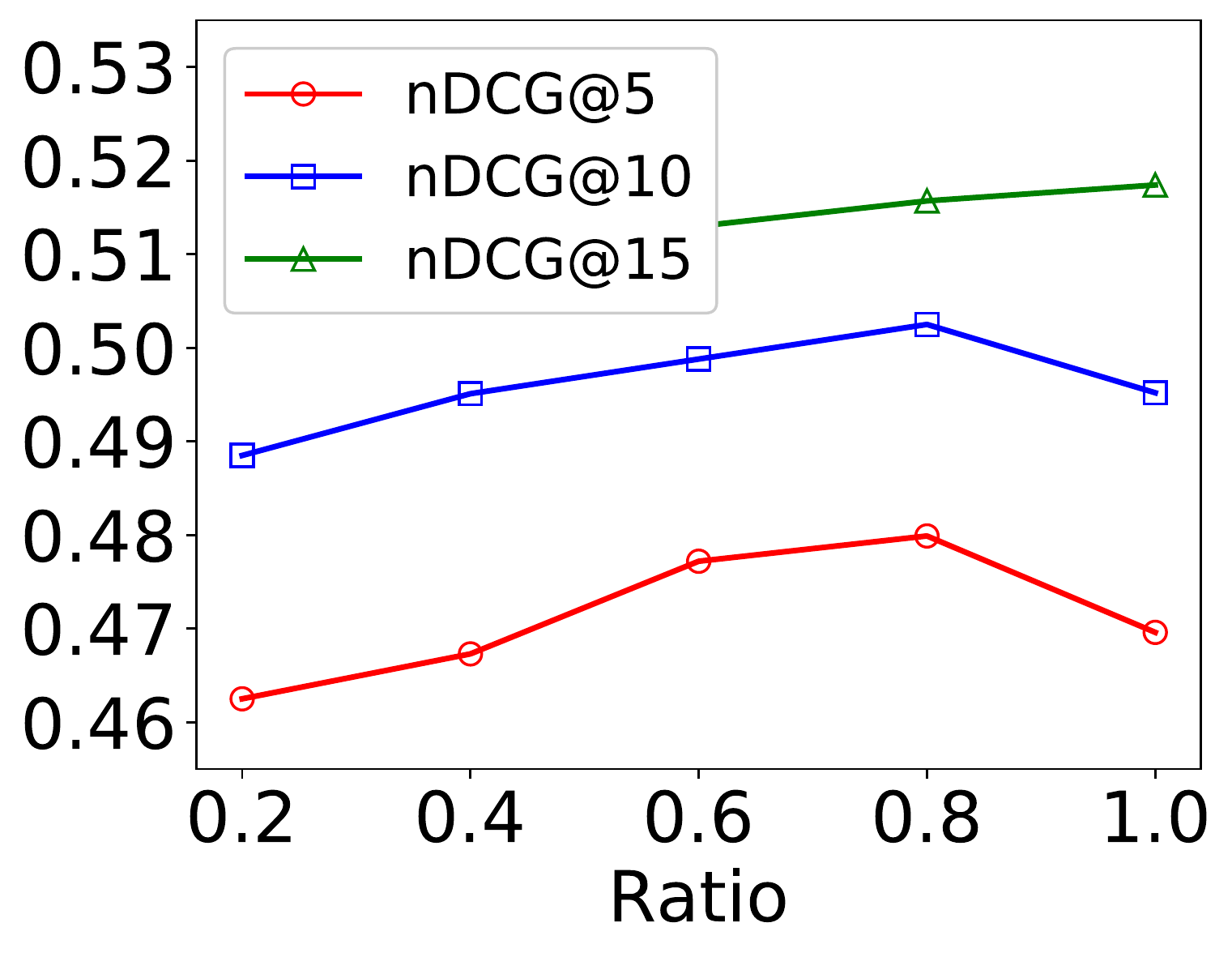}
		\caption{nDCG}
		\label{fig:ex_3_ndcg}
	\end{subfigure}
	\caption{Model performances with varying ratios of actions.}
	\label{fig:scalability}
\end{figure}

\begin{figure*}[t!]
	\centering
	\begin{subfigure}[b]{0.47\textwidth}
		\includegraphics[width=\textwidth]{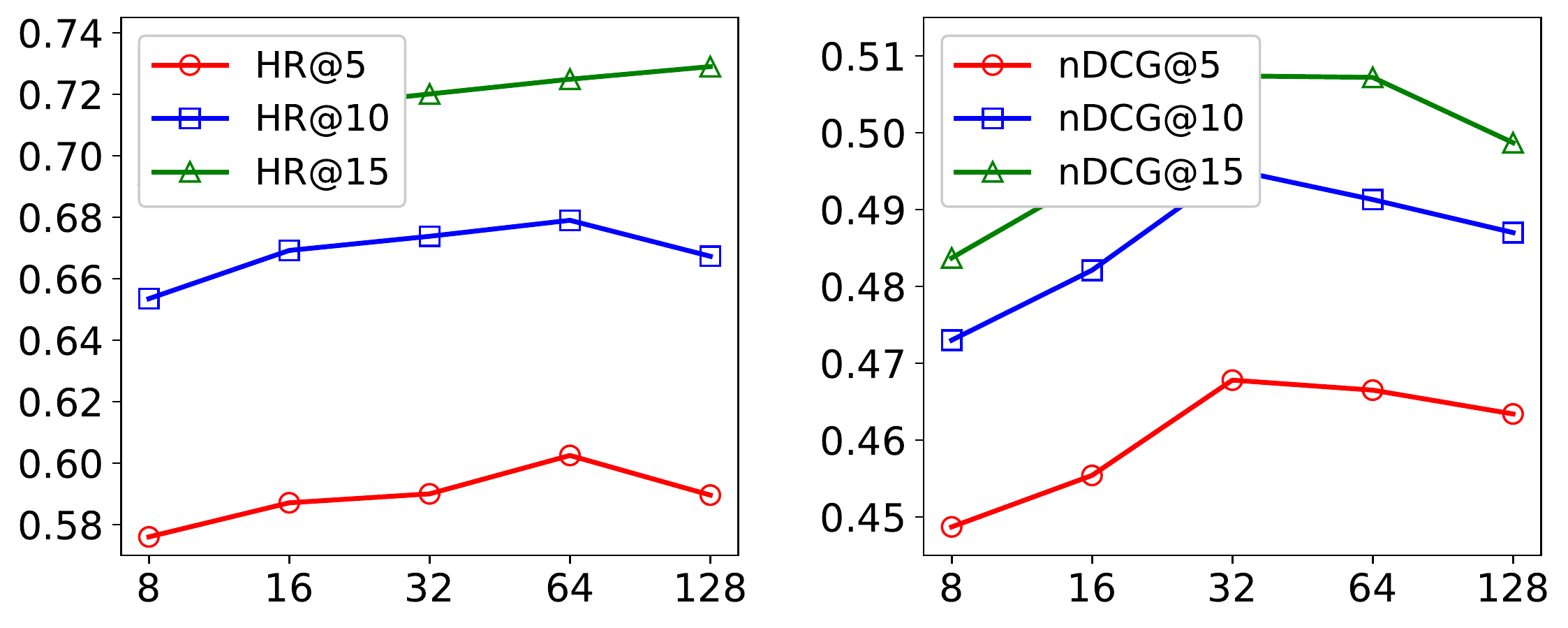}
		\caption{$d$}
		\label{fig:se_embed}
	\end{subfigure}
	\begin{subfigure}[b]{0.47\textwidth}
		\includegraphics[width=\textwidth]{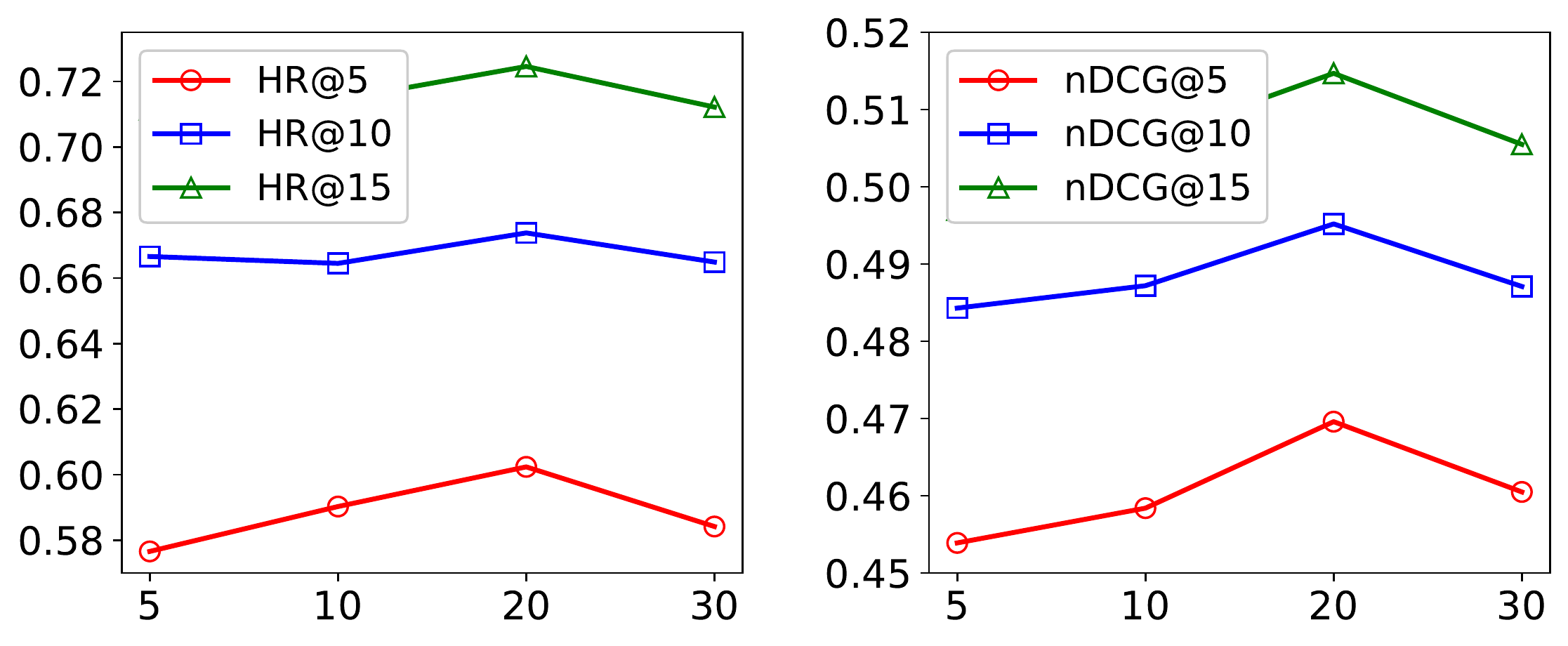}
		\caption{$m$}
		\label{fig:se_memory}
	\end{subfigure}
	\caption{Impact of hyper-parameters on model performances}
	\label{fig:sense_all}
\end{figure*}
The computational bottleneck of the proposed model when computing recommendation score lies in that we need to revisit the relation embedding between the target item and all the previously items of the user according to Equation~\eqref{eq:score}, which can be rather time consuming. Thus, we further validate whether our model can produce robust results if we only retain a partial relations with varying ratios ranging from 0.2 to 1.0 to make recommendation. As shown in Figure~\ref{fig:scalability}, we can see that the model performs quite robust against different ratios, with the largest difference being less than 2\%. It is also interesting to discover that using all user actions does not always produce the best results. For example, nDCG@5 and nDCG@10 achieve the best when only 80\% actions are available, which may be due to the fact that some relations may not indicate preferences and can be neglected. Thus, in the real scenarios, we only need to randomly sample a small number of user actions when making recommendations, which allows us to recommend more efficiently.

\subsection{Parameter Sensitivity}
In this section, we examine the sensitivity of two important parameters, \emph{i.e.}, the embedding size $d$ and the number of memory slices $m$. 

\textbf{Embedding size.} Figure~\ref{fig:se_embed} demonstrates the impact of embedding size on the results. It's easy to find that 64 is the best embedding size for HR while 32 is the best embedding size for nDCG. In addition, nDCG seems to be more sensitive to this parameter. We can see the performance of nDCG first increases along with $d$, and then begins to be stable when $d$ is larger than 32.

\textbf{The number of memory slices.} From the results shown in Figure~\ref{fig:se_memory}, we can find that both HR and nDCG achieve best at $m=20$. A possible reason is that when $m$ is relatively small, the memory attention layer lacks ability to capture multiple relations between items; while when $m$ is too big, the model might be overfitting.

In general, the performances remain relatively stable in all the settings, which shows the robustness of the method.

\begin{figure}[t!]
	\centering
	\begin{subfigure}[b]{0.21\textwidth}
		\includegraphics[width=\textwidth]{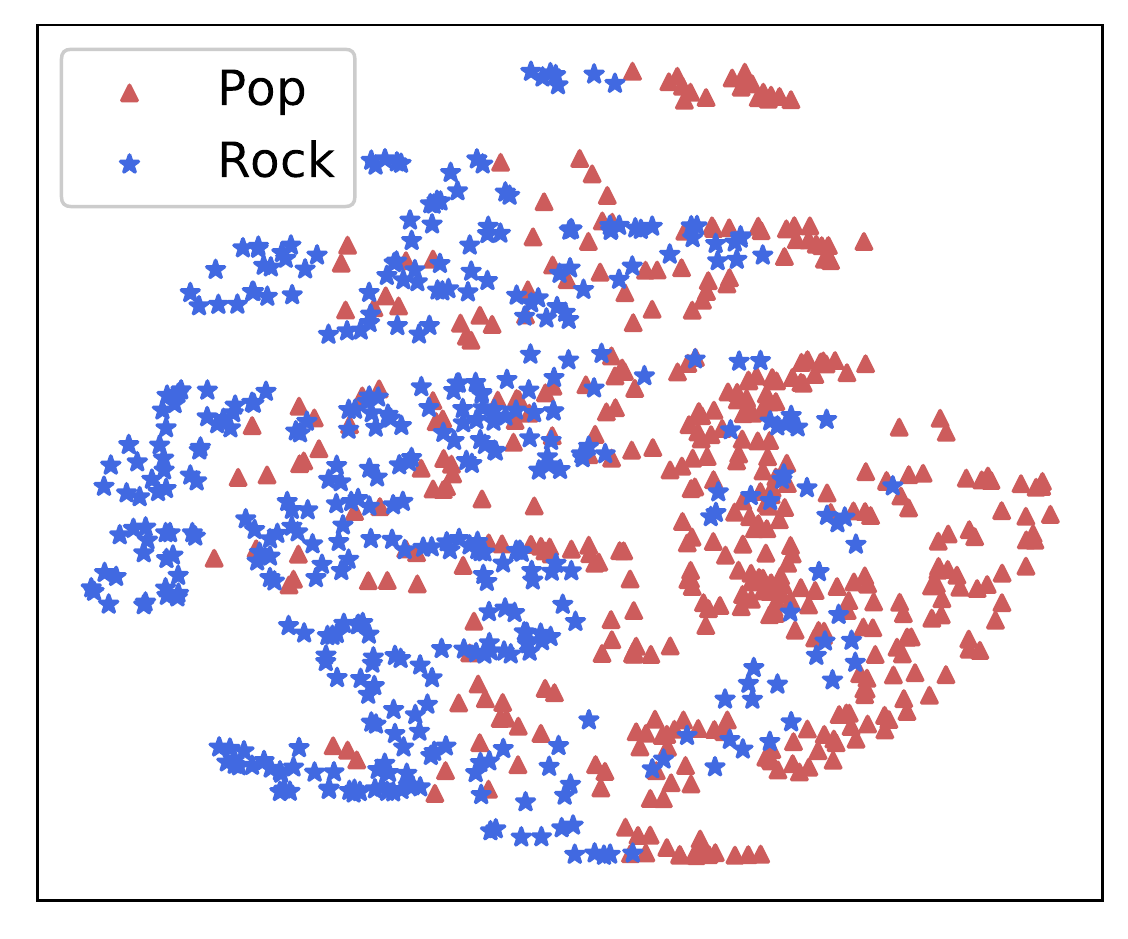}
		\label{fig:case_broad}
	\end{subfigure}
	\begin{subfigure}[b]{0.21\textwidth}
		\includegraphics[width=\textwidth]{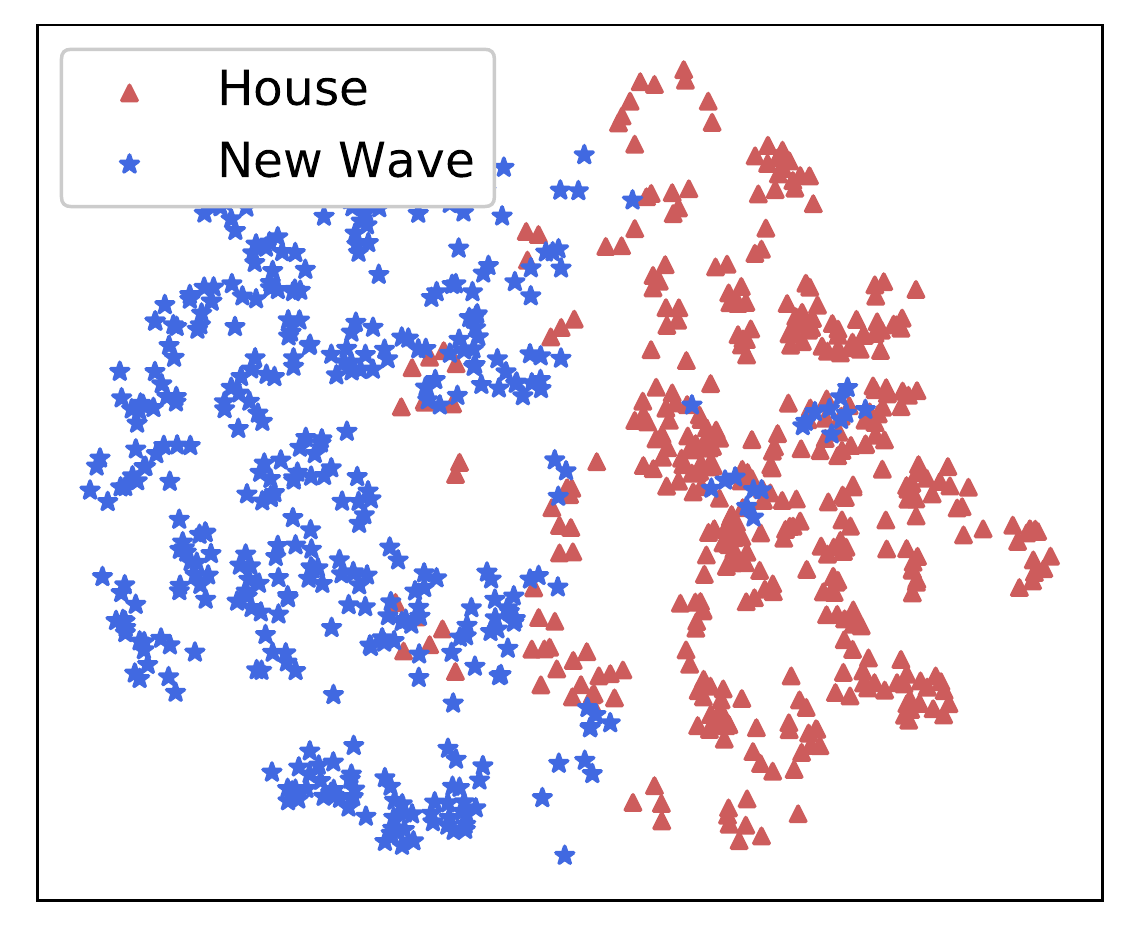}
		\label{fig:case_specific}
	\end{subfigure}
	\caption{A show case of learned item relations.}
	\label{fig:case_study}
\end{figure}

\subsection{Case Study}
We introduce auxiliary information such as item attributes to help us uncover what the model is learning, which is not provided during model training.
We firstly select several item pairs which share a certain genre, \emph{e.g.}, {\it Rock} and {\it Pop}, and then obtain their item relation embeddings from the model. For the sake of visualization, we exploit t-SNE~\cite{tsne} to project the derived relation embeddings to two-dimensional space as shown in Figure~\ref{fig:case_study}. Each group in this figure represents a special item relation. Firstly, as seen in Figure~\ref{fig:case_study}, the derived item relation embeddings for music with different genres are separated, showing that  REDA does capture fine-grained item relations in different aspects, and these aspects can be explained as several item attributes which are indeed not available during model training. 
In addition, we can discover that the relation embeddings of {\it Rock} and {\it Pop} music has more overlaps, while {\it House} and {\it New Wave} are clearly separated apart from each other. The underlying reason might be that the two genres {\it Rock} and {\it Pop} are both popular types music and many users would like both genres, and hence the item relations formed in these two groups may not be solely attributed to the genre.
In contrast, {\it House} and {\it New Wave} are relatively niche music and the item relations formed within each group are relatively pure and thus the embeddings can be well separated.

\begin{figure}[t!]
	\centering
	\includegraphics[width=0.3\textwidth]{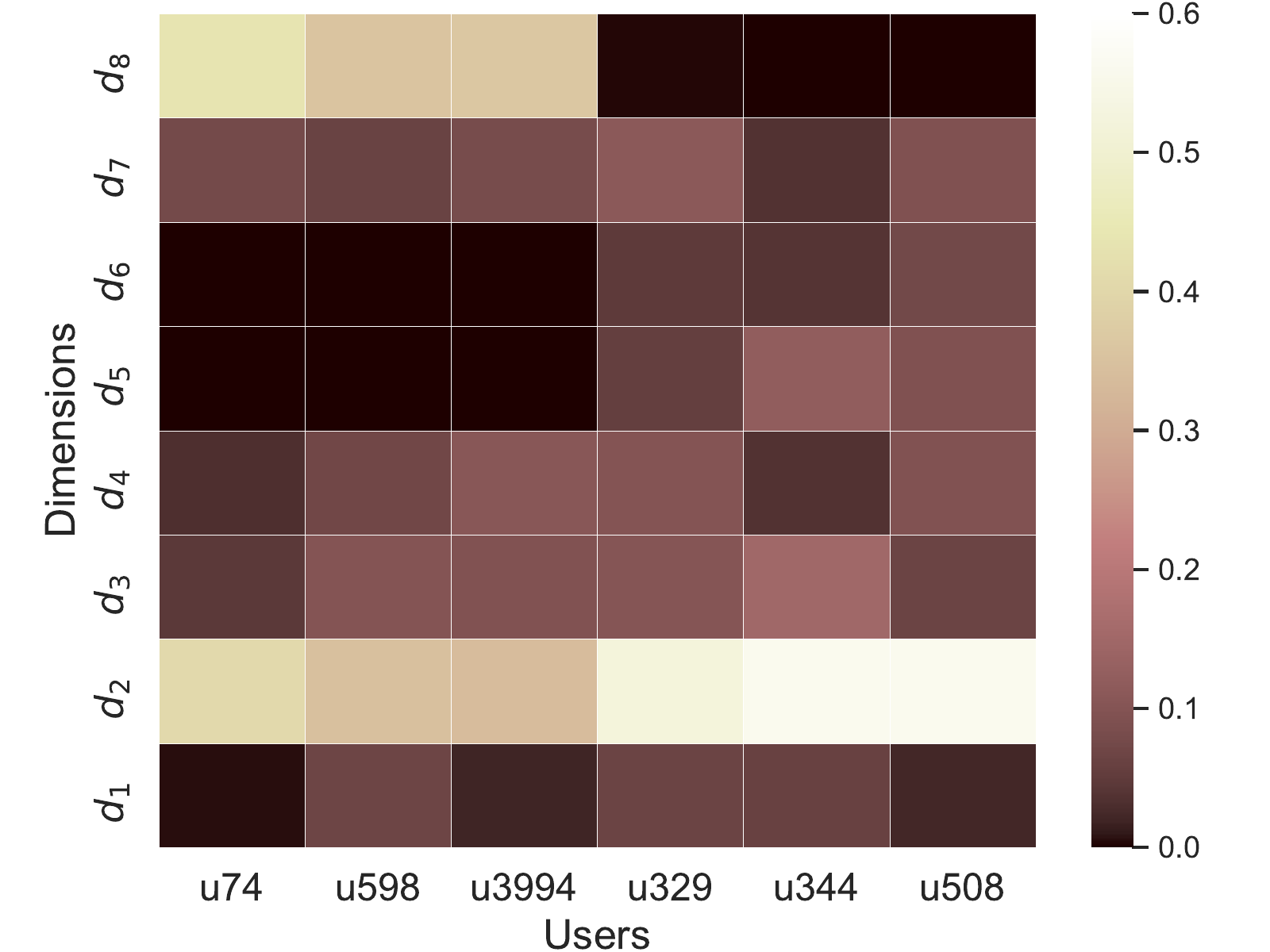}
	\caption{A show case of learned user representations.}
	\label{fig:user_resp}
\end{figure}

The meaning of learned user representations is also explored in this section. We first select several users who prefer music with a certain genre as an illustrative case. Then, we exploit softmax function to derive user representations aggregated from item relation embeddings for the sake of visualization. Specifically, {\it u}74, {\it u}598 and {\it u}3994 is keen on {\it Rock} while {\it u}329, {\it u}344 and {\it u}508 prefer {\it Hip-Hop} most. As shown in Figure~\ref{fig:user_resp}, we can discover that {\it Rock} users are mostly associated with a relatively uniform distribution on the dimension $d_2$ and $d_8$. While for {\it Hip-Hop} users, $d_2$ is the absolutely dominant dimension. 

To conclude, despite being only trained on implicit interactions arise from co-purchase records, the model REDA can still encode latent aspects of item pairs and can provide richer semantics than an overall similarity.

\section{Conclusion}
In this paper, we propose a personalized Item-based CF framework REDA to model the latent item relations for recommendation. By mapping the paired item into a relation space through a dual attention mechanism, REDA decomposes the overall similarity into multiple item relations. Meanwhile, a special embedding method namely relational user embedding is exploited to capture the fine-grained user preference and alleviate data sparsity problem simultaneously. In addition, a novel and efficient relation-wise optimization method is proposed to guide the learning of REDA. Experiments on four real-world datasets demonstrate the superiority of REDA, and the improvements increase as the datasets become sparser. The model robustness is also explored for faster recommendation with minimal reduction in the recommendation quality. Moreover, qualitative analysis further demonstrates that attribute information is actually encoded despite being only trained on implicit interactions. 

\clearpage

\bibliography{related_works}

\begin{thebibliography}{}

\bibitem[\protect\citeauthoryear{Bai \bgroup et al\mbox.\egroup
  }{2017}]{bai2017neural}
Bai, T.; Wen, J.-R.; Zhang, J.; and Zhao, W.~X.
\newblock 2017.
\newblock A neural collaborative filtering model with interaction-based
  neighborhood.
\newblock In {\em Proceedings of the 2017 ACM on Conference on Information and
  Knowledge Management},  1979--1982.
\newblock ACM.

\bibitem[\protect\citeauthoryear{Chen \bgroup et al\mbox.\egroup
  }{2018}]{chen2018sequential}
Chen, X.; Xu, H.; Zhang, Y.; Tang, J.; Cao, Y.; Qin, Z.; and Zha, H.
\newblock 2018.
\newblock Sequential recommendation with user memory networks.
\newblock In {\em Proceedings of the eleventh ACM international conference on
  web search and data mining},  108--116.
\newblock ACM.

\bibitem[\protect\citeauthoryear{Christakopoulou and
  Karypis}{2016}]{christakopoulou2016local}
Christakopoulou, E., and Karypis, G.
\newblock 2016.
\newblock Local item-item models for top-n recommendation.
\newblock In {\em Proceedings of the 10th ACM Conference on Recommender
  Systems},  67--74.
\newblock ACM.

\bibitem[\protect\citeauthoryear{Ebesu, Shen, and
  Fang}{2018}]{ebesu2018collaborative}
Ebesu, T.; Shen, B.; and Fang, Y.
\newblock 2018.
\newblock Collaborative memory network for recommendation systems.
\newblock In {\em The 41st International ACM SIGIR Conference on Research \&
  Development in Information Retrieval},  515--524.
\newblock ACM.

\bibitem[\protect\citeauthoryear{He and McAuley}{2016}]{he2016vbpr}
He, R., and McAuley, J.
\newblock 2016.
\newblock Vbpr: visual bayesian personalized ranking from implicit feedback.
\newblock In {\em Thirtieth AAAI Conference on Artificial Intelligence}.

\bibitem[\protect\citeauthoryear{He \bgroup et al\mbox.\egroup
  }{2017}]{he2017neural}
He, X.; Liao, L.; Zhang, H.; Nie, L.; Hu, X.; and Chua, T.-S.
\newblock 2017.
\newblock Neural collaborative filtering.
\newblock In {\em Proceedings of the 26th international conference on world
  wide web},  173--182.
\newblock International World Wide Web Conferences Steering Committee.

\bibitem[\protect\citeauthoryear{Hidasi \bgroup et al\mbox.\egroup
  }{2015}]{hidasi2015session}
Hidasi, B.; Karatzoglou, A.; Baltrunas, L.; and Tikk, D.
\newblock 2015.
\newblock Session-based recommendations with recurrent neural networks.
\newblock {\em arXiv preprint arXiv:1511.06939}.

\bibitem[\protect\citeauthoryear{Hsieh \bgroup et al\mbox.\egroup
  }{2017}]{hsieh2017collaborative}
Hsieh, C.-K.; Yang, L.; Cui, Y.; Lin, T.-Y.; Belongie, S.; and Estrin, D.
\newblock 2017.
\newblock Collaborative metric learning.
\newblock In {\em Proceedings of the 26th international conference on world
  wide web},  193--201.
\newblock International World Wide Web Conferences Steering Committee.

\bibitem[\protect\citeauthoryear{Kabbur, Ning, and
  Karypis}{2013}]{kabbur2013fism}
Kabbur, S.; Ning, X.; and Karypis, G.
\newblock 2013.
\newblock Fism: factored item similarity models for top-n recommender systems.
\newblock In {\em Proceedings of the 19th ACM SIGKDD international conference
  on Knowledge discovery and data mining},  659--667.
\newblock ACM.

\bibitem[\protect\citeauthoryear{Kim \bgroup et al\mbox.\egroup
  }{2016}]{kim2016convolutional}
Kim, D.; Park, C.; Oh, J.; Lee, S.; and Yu, H.
\newblock 2016.
\newblock Convolutional matrix factorization for document context-aware
  recommendation.
\newblock In {\em Proceedings of the 10th ACM Conference on Recommender
  Systems},  233--240.
\newblock ACM.

\bibitem[\protect\citeauthoryear{Kingma and Ba}{2014}]{kingma2014adam}
Kingma, D.~P., and Ba, J.
\newblock 2014.
\newblock Adam: A method for stochastic optimization.
\newblock {\em arXiv preprint arXiv:1412.6980}.

\bibitem[\protect\citeauthoryear{Kumar \bgroup et al\mbox.\egroup
  }{2016}]{kumar2016ask}
Kumar, A.; Irsoy, O.; Ondruska, P.; Iyyer, M.; Bradbury, J.; Gulrajani, I.;
  Zhong, V.; Paulus, R.; and Socher, R.
\newblock 2016.
\newblock Ask me anything: Dynamic memory networks for natural language
  processing.
\newblock In {\em International conference on machine learning},  1378--1387.

\bibitem[\protect\citeauthoryear{Liang \bgroup et al\mbox.\egroup
  }{2018}]{liang2018variational}
Liang, D.; Krishnan, R.~G.; Hoffman, M.~D.; and Jebara, T.
\newblock 2018.
\newblock Variational autoencoders for collaborative filtering.
\newblock In {\em Proceedings of the 2018 World Wide Web Conference},
  689--698.
\newblock International World Wide Web Conferences Steering Committee.

\bibitem[\protect\citeauthoryear{Liu \bgroup et al\mbox.\egroup
  }{2017}]{liu2017related}
Liu, D.~C.; Rogers, S.; Shiau, R.; Kislyuk, D.; Ma, K.~C.; Zhong, Z.; Liu, J.;
  and Jing, Y.
\newblock 2017.
\newblock Related pins at pinterest: The evolution of a real-world recommender
  system.
\newblock In {\em Proceedings of the 26th International Conference on World
  Wide Web Companion},  583--592.
\newblock International World Wide Web Conferences Steering Committee.

\bibitem[\protect\citeauthoryear{McAuley \bgroup et al\mbox.\egroup
  }{2015}]{mcauley2015image}
McAuley, J.; Targett, C.; Shi, Q.; and Van Den~Hengel, A.
\newblock 2015.
\newblock Image-based recommendations on styles and substitutes.
\newblock  43--52.
\newblock ACM.

\bibitem[\protect\citeauthoryear{Miller \bgroup et al\mbox.\egroup
  }{2016}]{miller2016key}
Miller, A.; Fisch, A.; Dodge, J.; Karimi, A.-H.; Bordes, A.; and Weston, J.
\newblock 2016.
\newblock Key-value memory networks for directly reading documents.
\newblock {\em arXiv preprint arXiv:1606.03126}.

\bibitem[\protect\citeauthoryear{Ning and Karypis}{2011}]{ning2011slim}
Ning, X., and Karypis, G.
\newblock 2011.
\newblock Slim: Sparse linear methods for top-n recommender systems.
\newblock In {\em 2011 11th IEEE International Conference on Data Mining},
  497--506.
\newblock IEEE.

\bibitem[\protect\citeauthoryear{Paatero and
  Tapper}{2010}]{Paatero2010Positive}
Paatero, P., and Tapper, U.
\newblock 2010.
\newblock Positive matrix factorization: A non-negative factor model with
  optimal utilization of error estimates of data values.
\newblock {\em Environmetrics} 5(2):111--126.

\bibitem[\protect\citeauthoryear{Rendle \bgroup et al\mbox.\egroup
  }{2009}]{rendle2009bpr}
Rendle, S.; Freudenthaler, C.; Gantner, Z.; and Schmidt-Thieme, L.
\newblock 2009.
\newblock Bpr: Bayesian personalized ranking from implicit feedback.
\newblock In {\em Proceedings of the twenty-fifth conference on uncertainty in
  artificial intelligence},  452--461.
\newblock AUAI Press.

\bibitem[\protect\citeauthoryear{Rendle}{2010}]{rendle2010factorization}
Rendle, S.
\newblock 2010.
\newblock Factorization machines.
\newblock In {\em 2010 IEEE International Conference on Data Mining},
  995--1000.
\newblock IEEE.

\bibitem[\protect\citeauthoryear{Shang \bgroup et al\mbox.\egroup
  }{2019}]{shang2019gamenet}
Shang, J.; Xiao, C.; Ma, T.; Li, H.; and Sun, J.
\newblock 2019.
\newblock Gamenet: Graph augmented memory networks for recommending medication
  combination.
\newblock In {\em Proceedings of the AAAI Conference on Artificial
  Intelligence}, volume~33,  1126--1133.

\bibitem[\protect\citeauthoryear{Smith and Linden}{2017}]{smith2017two}
Smith, B., and Linden, G.
\newblock 2017.
\newblock Two decades of recommender systems at amazon. com.
\newblock {\em Ieee internet computing} 21(3):12--18.

\bibitem[\protect\citeauthoryear{Sukhbaatar \bgroup et al\mbox.\egroup
  }{2015}]{sukhbaatar2015end}
Sukhbaatar, S.; Weston, J.; Fergus, R.; et~al.
\newblock 2015.
\newblock End-to-end memory networks.
\newblock In {\em Advances in neural information processing systems},
  2440--2448.

\bibitem[\protect\citeauthoryear{Tay, Anh~Tuan, and Hui}{2018}]{tay2018latent}
Tay, Y.; Anh~Tuan, L.; and Hui, S.~C.
\newblock 2018.
\newblock Latent relational metric learning via memory-based attention for
  collaborative ranking.
\newblock In {\em Proceedings of the 2018 World Wide Web Conference},
  729--739.
\newblock International World Wide Web Conferences Steering Committee.

\bibitem[\protect\citeauthoryear{van~der Maaten and Hinton}{2008}]{tsne}
van~der Maaten, L., and Hinton, G.~E.
\newblock 2008.
\newblock Visualizing high-dimensional data using t-sne.
\newblock {\em JMLR} 9:2579--2605.

\bibitem[\protect\citeauthoryear{Wan and McAuley}{2018}]{wan2018item}
Wan, M., and McAuley, J.
\newblock 2018.
\newblock Item recommendation on monotonic behavior chains.
\newblock In {\em Proceedings of the 12th ACM Conference on Recommender
  Systems},  86--94.
\newblock ACM.

\bibitem[\protect\citeauthoryear{Wang \bgroup et al\mbox.\egroup
  }{2017}]{wang2017deep}
Wang, R.; Fu, B.; Fu, G.; and Wang, M.
\newblock 2017.
\newblock Deep \& cross network for ad click predictions.
\newblock In {\em Proceedings of the ADKDD'17}, ~12.
\newblock ACM.

\bibitem[\protect\citeauthoryear{Wu \bgroup et al\mbox.\egroup
  }{2017}]{wu2017recurrent}
Wu, C.-Y.; Ahmed, A.; Beutel, A.; Smola, A.~J.; and Jing, H.
\newblock 2017.
\newblock Recurrent recommender networks.
\newblock In {\em Proceedings of the tenth ACM international conference on web
  search and data mining},  495--503.
\newblock ACM.

\end{thebibliography}
\bibliographystyle{aaai}

\end{document}